\documentclass[12pt,psfig]{article}
\usepackage{natbib, psfig}
\def\reference{\parskip 0pt\par\noindent\hangindent 0.5 truecm}

\def \etal{et al.}
\def \cup{Cambridge U. Press}
\def \cupadr{Cambridge}
\def \pup{Princeton U. Press}
\def \pupadr{Princeton, U.S.A.}

\def \ajp{Am. J. Phys.}
\def \apj{ApJ}

\def \mnras{MNRAS}

\def \prd{Phys. Rev. D}

\def \qjras{QJRAS}

\newcommand{\beq}{\begin{equation}}
\newcommand{\eeq}{\end{equation}}
\newcommand{\bea}{\begin{eqnarray}}
\newcommand{\eea}{\end{eqnarray}}
\newcommand{\bctr}{\begin{center}}
\newcommand{\ectr}{\end{center}}
\newcommand{\fns}{\footnotesize}
\newcommand{\om}{\Omega_{\rm M}}
\newcommand{\oll}{\Omega_{\Lambda}}
\newcommand{\omol}{(\Omega_{\rm M},\Omega_{\Lambda})}
\newcommand{\vrec}{v_{\rm rec}}
\newcommand{\vp}{v_{\rm pec}}

\newcommand{\lsim}{\mbox{$\:\stackrel{<}{_{\sim}}\:$} }
\newcommand{\gsim}{\mbox{$\:\stackrel{>}{_{\sim}}\:$} }
\newcommand{\incite}[1]{[App.~B: {#1}]} 
\newcommand{\inciteFirst}[1]{[Appendix~B: {#1}]} 
\newcommand{\xcite}[2]{[#1] {#2}\newline} 
\begin{document}
%
%
\title{Expanding Confusion:\\ common misconceptions of cosmological horizons\\
 and the superluminal expansion of the universe}
%


\author{Tamara M. Davis $^{1}$ \and 
 Charles H. Lineweaver $^{1}$ 
} 

\date{}
\maketitle

{\center
$^1$ University of New South Wales, Sydney, Australia, 2052\\tamarad@phys.unsw.edu.au\\charley@bat.phys.unsw.edu.au\\[3mm]
}

%
\begin{abstract}
We use standard general relativity to illustrate and clarify several common misconceptions about the expansion of the universe.  To show the abundance of these misconceptions we cite numerous misleading, or easily misinterpreted, statements in the literature.
In the context of the new standard $\Lambda$CDM cosmology we point out 
confusions regarding the particle horizon, the event horizon, the ``observable universe'' and the Hubble sphere (distance at which recession velocity $= c$).  We show that we can observe galaxies that have, and always have had, recession velocities greater than the speed of light.  We explain why this does not violate special relativity and we link these concepts to observational tests.   Attempts to restrict recession velocities to less than the speed of light require a special relativistic interpretation of cosmological redshifts. 
We analyze apparent magnitudes of supernovae and observationally rule out the special relativistic Doppler interpretation of cosmological redshifts at a confidence level of $23 \sigma$. 
\end{abstract}


\bigskip

\section{Introduction}
The general relativistic (GR) interpretation of the redshifts of distant galaxies, as the expansion of the universe, 
is widely accepted. However this interpretation leads to several concepts that are widely misunderstood.
Since the expansion of the universe is the basis of the big bang model, these misunderstandings
are fundamental.
Popular science books written by astrophysicists, astrophysics textbooks and to some extent professional astronomical
literature addressing the expansion of the Universe, contain  misleading, or easily misinterpreted, statements concerning recession 
velocities, horizons and the ``observable universe''. 
Probably the most common misconceptions surround the expansion of the Universe at distances beyond which Hubble's law ($v_{\rm rec} = H D$: recession velocity = Hubble's constant $\times$ distance) predicts recession velocities faster than the speed of light~\inciteFirst{1--8}, 
despite efforts to clarify the issue (Murdoch 1977, Harrison 1981, Silverman 1986, Stuckey 1992, Ellis \& Rothman 1993, Harrison 1993, Kiang 1997, Davis \& Lineweaver 2000, Kiang 2001, Gudmundsson and Bj\"ornsson 2002). 
Misconceptions include misleading comments about the observability of objects receding faster than light~\incite{9--13}. 
Related, but more subtle confusions can be found surrounding cosmological event horizons~\incite{14--15}. 
The concept of the expansion of the universe is so fundamental to our understanding of cosmology 
and the misconceptions so abundant that it is important to clarify these issues and make the connection with observational tests as explicit as possible. 
In Section~\ref{sect:fig1} we review and illustrate the standard general relativistic description of the expanding universe using spacetime diagrams and we provide a mathematical summary in Appendix~\ref{sect:math}.  On the basis of this description, in Section~\ref{sect:misconceptions} we point out and clarify common misconceptions about superluminal recession velocities and horizons.  Examples of misconceptions, or easily misinterpreted statements, occurring in the literature are given in Appendix~\ref{sect:quotes}.  Finally, in Section~\ref{sect:data} we provide explicit observational tests demonstrating that attempts to apply special relativistic concepts to the Universe are in conflict with observations. 

\begin{figure*}
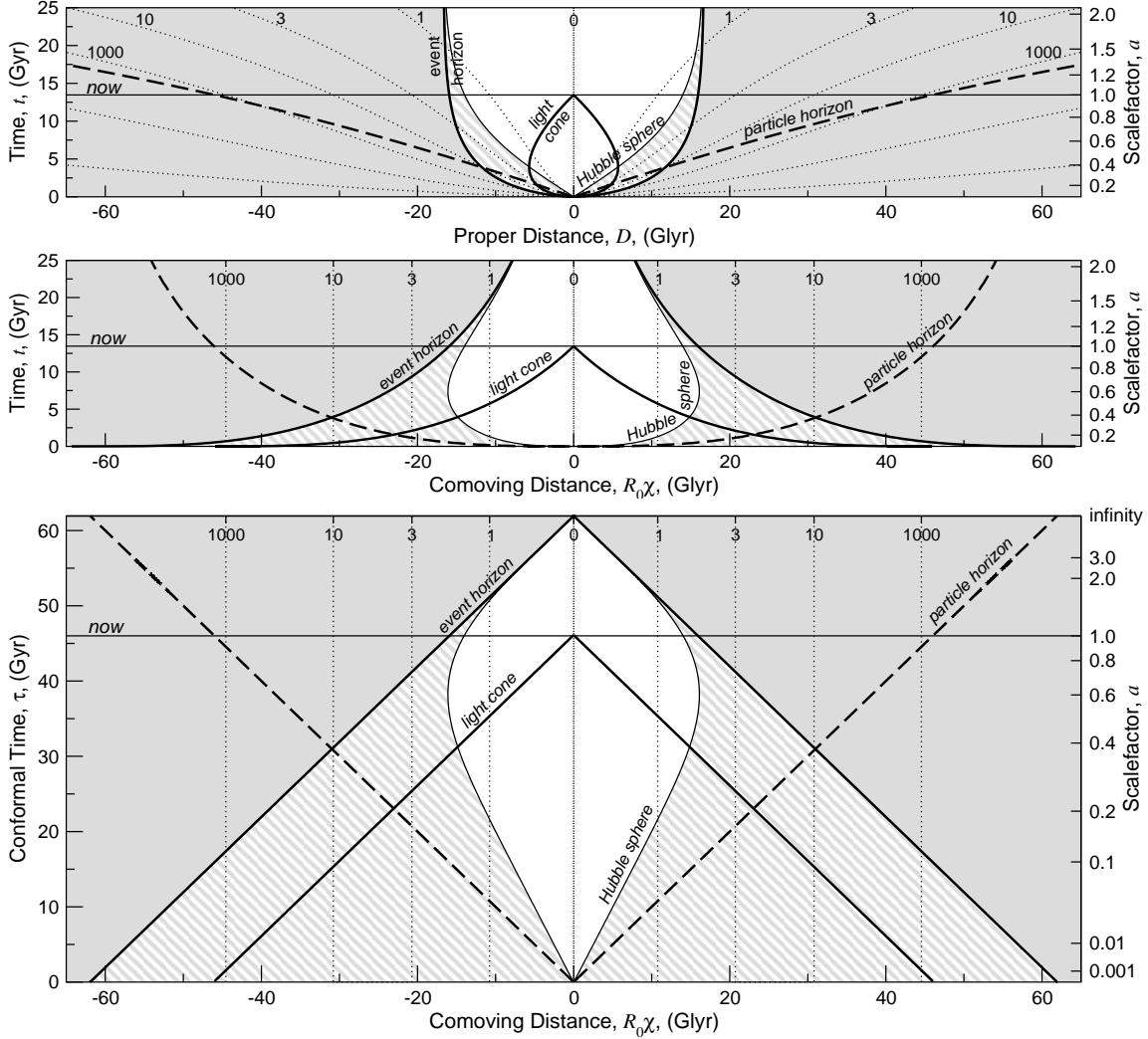
\bctr
\psfig{file=fig1a.eps,width=152mm}
\psfig{file=fig1b.eps,width=152mm}
\psfig{file=fig1c.eps,width=152mm}
\vspace{-3mm}
\renewcommand\baselinestretch{1.0}
\caption{\fns Spacetime diagrams showing the main features of the general relativistic description of the expansion of the universe for the $\omol=(0.3,0.7)$ model with $H_0= 70\,km\; s^{-1} Mpc^{-1}$.  Dotted lines show the worldlines of comoving objects.  We are the central vertical worldline.  The current redshifts of the comoving galaxies shown appear labeled on each comoving worldline.  The normalized scalefactor, $a=R/R_0$, is drawn as an alternate vertical axis.  
All events that we currently observe are on our past light cone (with apex at $t={\rm now}$).
All comoving objects beyond the Hubble sphere (thin solid line) are receding faster than the speed of light.
%
Top panel (proper distance): The speed of photons relative to us (the slope of the light cone) is not constant, but is rather $v_{\rm rec}-c$.  
Photons we receive that were emitted by objects beyond the Hubble sphere were initially receding from us (outward sloping lightcone at $t\lsim 5$ Gyr).  Only when they passed from the region of superluminal recession $v_{\rm rec}>c$ (gray crosshatching) to the region of subluminal recession (no shading) can the photons approach us.
More detail about early times and the horizons is visible in comoving coordinates (middle panel) and conformal coordinates (lower panel).  Our past light cone in comoving coordinates appears to approach the horizontal ($t=0$) axis asymptotically.  However it is clear in the lower panel that the past light cone at $t=0$ only reaches a finite distance: about 46 Glyr, the current distance to the particle horizon.  Currently observable light that has been travelling towards us since the beginning of the universe, was emitted from comoving positions that are now $46$ Glyr from us.  
The distance to the particle horizon as a function of time is represented by the dashed line.  
Our event horizon is our past light cone at the end of time, $t=\infty$ in this case.  It asymptotically approaches $\chi=0$ as $t\rightarrow \infty$.  
The vertical axis of the lower panel shows conformal time.  An infinite proper time is transformed into a finite conformal time so this diagram is complete on the vertical axis.
The aspect ratio of $\sim 3/1$ in the top two panels represents the ratio between the radius of the observable universe and the age of the universe, 46 Glyr/13.5 Gyr.
}
\label{fig:dist}\ectr
\end{figure*}

\section{Standard general relativistic description of expansion}
\label{sect:fig1}
The results in this paper are based on the standard general relativistic description of an expanding homogeneous, isotropic universe (Appendix~\ref{sect:math}).  Here we briefly summarize the main features of the GR description, about which misconceptions often arise.  On the spacetime diagrams in Fig.~\ref{fig:dist} we demonstrate these features for the observationally favoured $\Lambda$CDM concordance model: $(\Omega_{\rm M},\Omega_{\Lambda})= (0.3,0.7)$ with $H_0=70\,kms^{-1}Mpc^{-1}$ (Bennett \etal~2003, to one significant figure). 
The three spacetime diagrams in Fig.~\ref{fig:dist} plot, from top to bottom, time versus proper distance $D$, time versus comoving distance $R_0\chi$, and conformal time $\tau$ versus comoving distance.  They show the relationship between comoving objects, our past light cone, the Hubble sphere and cosmological horizons. 

Two types of horizon are shown in Fig.~\ref{fig:dist}.  The particle horizon is the distance light can have travelled from $t=0$ to a given time $t$ (Eq.~\ref{eq:chipht}), whereas the event horizon is the distance light can travel from a given time $t$ to $t=\infty$ (Eq.~\ref{eq:eventhorizont}).  Using Hubble's law ($v_{\rm rec}=HD$), the Hubble sphere is defined to be the distance beyond which the recession velocity exceeds the speed of light, $D_{\rm HS}=c/H$.  As we will see, the Hubble sphere is not an horizon.  Redshift does not go to infinity for objects on our Hubble sphere (in general) and for many cosmological models we can see beyond it.

In the $\Lambda$CDM concordance model all objects with redshift greater than $z\sim1.46$ are receding faster than the speed of light.  
This does not contradict SR because the motion is not in any observer's inertial frame.  No observer ever overtakes a light beam and all observers measure light locally to be travelling at $c$.  Hubble's law is derived directly from the Robertson-Walker metric (Eq.~\ref{eq:frwmetric}), and is valid for all distances in any homogeneous, expanding universe.   

The teardrop shape of our past light cone in the top panel of Fig.~\ref{fig:dist} shows why we can observe objects that are receding superluminally.  Light that superluminally receding objects emit propagates towards us with a local peculiar velocity of $c$, but since the recession velocity at that distance is greater than $c$, the total velocity of the light is away from us (Eq.~\ref{eq:vtot}).    However, since the radius of the Hubble sphere increases with time, some photons that were initially in a superluminally receding region later find themselves in a subluminally receding region.  They can therefore approach us and eventually reach us.    The objects that emitted the photons however, have moved to larger distances and so are still receding superluminally.  Thus we can observe objects that are receding faster than the speed of light (see Section~\ref{sect:notobserve} for more detail). 

Our past light cone approaches the cosmological event horizon as $t_0 \rightarrow \infty$ (Eqs.~\ref{eq:lightconet} and~\ref{eq:eventhorizont}).
Most observationally viable cosmological models have event horizons and in the $\Lambda$CDM model of Fig.~\ref{fig:dist}, galaxies with redshift $z\sim1.8$ are currently crossing our event horizon.  These are the most distant objects from which we will ever be able to receive information about the present day. 
The particle horizon marks the size of our observable universe.  It is the distance to the most distant object we can see at any particular time.
The particle horizon can be larger than the event horizon because, although we cannot see {\em events} that occur beyond our event horizon, we can still see many  galaxies that are beyond our current event horizon by light they emitted long ago.  

In the GR description of the expansion of the Universe redshifts do not relate to velocities according to any SR expectations.  We do not observe objects on the Hubble sphere (that recede at the speed of light) to have an infinite redshift (solve Eq.~\ref{eq:chiz} for $z$ using $\chi = c/\dot{R}$).  
Instead photons we receive that have infinite redshift were emitted by objects on our particle horizon. 
In addition, all galaxies become increasingly redshifted as we watch them approach the cosmological {\em event} horizon ($z\rightarrow\infty$ as $t\rightarrow\infty$).  As the end of the universe approaches, all objects that are not gravitationally bound to us will be redshifted out of detectability.

Since this paper deals frequently with recession velocities and the expansion of the Universe the observational status of these concepts is important and is discussed in Sections~\ref{sect:data} and~\ref{sect:discussion}.

\section{Misconceptions}
\label{sect:misconceptions}
\subsection{Misconception \#1: Recession velocities cannot exceed the speed of light}

A common misconception is that the expansion of the Universe cannot be faster than the speed of light. 
Since Hubble's law predicts superluminal recession at large distances ($D>c/H$) it is sometimes stated that Hubble's law needs special relativistic corrections when the recession velocity approaches the speed of light~\incite{6--7}. 
However, it is well-accepted that general relativity, not special relativity, is necessary to describe cosmological observations.  Supernovae surveys calculating cosmological parameters, galaxy-redshift surveys and cosmic microwave background anisotropy tests, all use general relativity to explain their observations.  When observables are calculated using special relativity, contradictions with observations quickly arise (Section~\ref{sect:sr}).  
Moreover, we know there is no contradiction with special relativity when faster than light motion occurs {\em outside 
the observer's inertial frame}.  General relativity was specifically derived to be able to predict motion when global inertial frames were not available.   
Galaxies that are receding from us superluminally are at rest locally (their peculiar velocity, $v_{\rm pec}=0$) and motion in their local inertial frames remains well described by special relativity.
They are in no sense catching up with photons ($v_{\rm pec}=c$).  Rather, the galaxies and the photons are both receding from us at recession velocities greater than the speed of light. 

In special relativity, redshifts arise directly from velocities.  It was this idea that led Hubble in 1929 to convert the redshifts of the ``nebulae'' he observed into velocities, and predict the expansion of the universe with the linear velocity-distance law that now bears his name.  
The general relativistic interpretation of the expansion interprets cosmological redshifts as an indication of velocity since the proper distance between comoving objects increases. 
However, the velocity is due to the rate of expansion of space, not movement through space, and therefore cannot be calculated with the special relativistic Doppler shift formula.  Hubble \& Humason's calculation of velocity therefore should not be given special relativistic corrections at high redshift, contrary to their suggestion~\incite{16}. 

The general relativistic and special relativistic relations between velocity and cosmological redshift are (e.g. Davis \& Lineweaver, 2001): 
\bea
{\rm GR}&\;\; v_{\rm rec}(t,z)  =& \frac{c}{R_0}\;\dot{R}(t)\int^{z}_{0}\frac{dz^\prime}{H(z^\prime)},  \label{eq:vGR}\\
                               & & \nonumber\\ 
{\rm SR} &\;\;\;\;\vp (z)       =& c \: \frac{(1+ z)^2-1}{(1 + z)^2+1}. \label{eq:vSR} 
\eea
These velocities are measured with respect to the comoving observer who observes the receding object to have redshift, $z$.
The GR description is written explicitly as a function of time because when we observe an object with redshift, $z$, we must specify the epoch at which we wish to calculate its recession velocity.  For example, setting $t= t_{o}$ yields the recession velocity today of the object that emitted the observed photons at $t_{\rm em}$. Setting $t=t_{\rm em}$ yields the recession velocity at the time the photons were emitted (see Eqs.~\ref{eq:vchi} \&~\ref{eq:chiz}).  The changing recession velocity of a comoving object is reflected in the changing slope of its worldline in the top panel of Fig.~\ref{fig:dist}.
There is no such time dependence in the SR relation.

\begin{figure}\bctr
\psfig{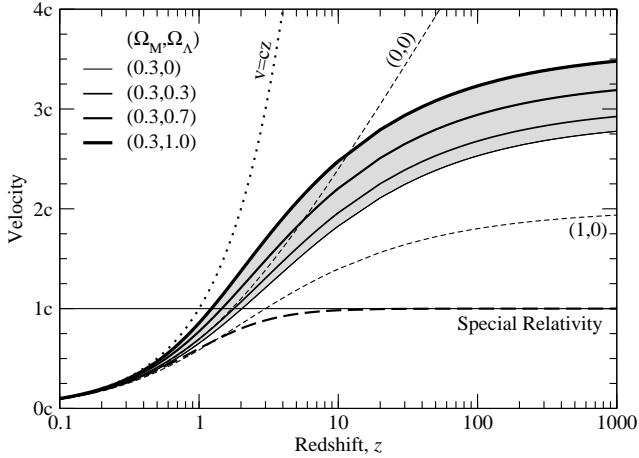}
\caption{Velocity as a function of redshift under various assumptions.  The linear approximation, $v=cz$, is the low redshift approximation of both the GR and SR results.  The SR result is calculated using Eq.~\ref{eq:vSR} while the GR result uses Eq.~\ref{eq:vGR} and quotes the recession velocity at the present day, i.e., uses $\dot{R}(t)=\dot{R}_0$. 
The solid dark lines and gray shading show a range of Fridemann-Robertson-Walker (FRW) models as labeled in the legend.  These include the observationally favoured cosmological model $\omol=(0.3,0.7)$.  The recession velocity of all galaxies with $z\gsim 1.5$ exceeds the speed of light in all viable cosmological models.  Observations now routinely probe regions that are receding faster than the speed of light.}
\label{fig:vz}\ectr
\end{figure}

Despite the fact that special relativity incorrectly describes cosmological redshifts it has been used for decades to convert cosmological redshifts into velocity because the special relativistic Doppler shift formula (Eq.~\ref{eq:vSR}), shares the same low redshift approximation, $v=cz$, as Hubble's Law (Fig.~\ref{fig:vz}).   
It has only been in the last decade that routine observations have been deep enough that the distinction makes a difference. 
Figure~\ref{fig:vz} shows a snapshot at $t_0$ of the GR velocity-redshift relation for various models as well as the SR velocity-redshift relation and their common low redshift approximation, $v=cz$.  Recession velocities exceed the speed of light in all viable cosmological models for objects with redshifts greater than $z\sim 1.5$.
At higher redshifts special relativistic ``corrections'' can be more incorrect than the simple linear approximation (Fig.~\ref{fig:mag-z}).

Some of the most common misleading applications of relativity arise from the misconception that nothing can recede faster than the speed of light.  These include texts asking students to calculate the velocity of a high redshift receding galaxy using the special relativistic Doppler shift equation~\incite{17--21}, 
as well as the comments that galaxies recede from us at speeds ``approaching the speed of light''~\incite{4--5, 8}, 
 or quasars are receding at a certain percentage of the speed of light\footnote{Redshifts are usually converted into velocities using $v=cz$, which is a good approximation for $z\lsim 0.3$ (see Fig.~\ref{fig:vz}) but inappropriate for today's high redshift measurements.  When a ``correction'' is made for high redshifts, the formula used is almost invariably the inappropriate special relativistic Doppler shift equation (Eq.~\ref{eq:vSR}).}~\incite{3, 18--21}. 

Although velocities of distant galaxies are in principle observable, the set of synchronized comoving observers required to measure proper distance (Weinberg, 1972, p.~415; Rindler, 1977, p.~218) 
 is not practical.  Instead, more direct observables such as the redshifts of standard candles can be used to observationally rule out the special relativistic interpretation of cosmological redshifts (Section~\ref{sect:data}).

\vspace{1cm}
\subsection{Misconception \#2: Inflation causes superluminal expansion of the universe but the normal expansion of the universe does not}
\label{sect:inflationissuperluminal}

Inflation is sometimes described as ``superluminal expansion''~\incite{22--23}. 
This is misleading because {\em any} expansion described by Hubble's law has superluminal expansion for sufficiently distant objects.  Even during inflation, objects within the Hubble sphere ($D<c/H$) recede at less than  the speed of light, while objects beyond the Hubble sphere ($D>c/H$) recede faster than the speed of light.  This is identical to the situation during non-inflationary expansion, except the Hubble constant during inflation was much larger than subsequent values. Thus the distance to the Hubble sphere was much smaller.  
During inflation the proper distance to the Hubble sphere stays 
constant and is coincident with the event horizon -- this is also identical to the asymptotic behaviour of an eternally expanding universe with a cosmological constant $\oll>0$ (Fig. \ref{fig:dist}, top panel).  

The oft-mentioned concept of structures ``leaving the horizon'' during the inflationary period refers to  structures once smaller than the Hubble sphere becoming larger than the Hubble sphere.  If the exponentially expanding regime, $R=R_0e^{Ht}$, were extended to the end of time, the Hubble sphere would be the event horizon.  However, in the context of inflation the Hubble sphere is not a true event horizon because structures that have crossed the horizon can ``reenter the horizon'' after inflation stops.  The horizon they ``reenter'' is the revised event horizon determined by how far light can travel in a FRW universe without inflation.  

It would be more appropriate to describe inflation as superluminal expansion if all distances down to the Planck length, $l_{\rm pl}\sim 10^{-35}m$, were receding faster than the speed of light.  Solving $D_{\rm H}=c/H = l_{\rm pl}$ gives $H = 10^{43}s^{-1}$ (inverse Planck time) which is equivalent to $H= 10^{62}kms^{-1}Mpc^{-1}$.   If Hubble's constant during inflation exceeded this value it would justify describing inflation as ``superluminal expansion''. 


\subsection{Misconception \#3: Galaxies with recession velocities exceeding the speed of light exist but we cannot see them}
\label{sect:notobserve}

Amongst those who acknowledge that recession velocities can exceed the speed of light, the claim is sometimes made that objects with recession velocities faster than the speed of light are not observable~\incite{9--13}. 
We have seen that the speed of photons propagating towards us (the slope of our past light cone in the upper panel of Fig.~\ref{fig:dist}) is not constant, but is rather $v_{\rm rec}-c$.  Therefore light that is beyond the Hubble sphere has a total velocity away from us.  How is it then that we can ever see this light?  Although the photons are in the superluminal region and therefore recede from us (in proper distance), the Hubble sphere also recedes.  In decelerating universes $H$ decreases as $\dot{a}$ decreases (causing the Hubble sphere to recede). In accelerating universes $H$ also tends to decrease since $\dot{a}$ increases more slowly than $a$.   As long as the Hubble sphere recedes faster than the photons immediately outside it, $\dot{D}_{\rm H}>v_{\rm rec}-c$, the photons end up in a subluminal region and approach us\footnote{The behaviour of the Hubble sphere is model dependent.    The Hubble sphere recedes as long as the deceleration parameter $q=-\ddot{R}R/\dot{R}^2>-1$.  In some closed eternally accelerating universes (specifically $\om +\oll>1$ and $\oll>0$) the deceleration parameter can be less than minus one in which case we see faster-than-exponential expansion and some subluminally expanding regions can be beyond the event horizon (light that was initially in subluminal regions can end up in superluminal regions and never reach us).  Exponential expansion, such as that found in inflation, has $q=-1$.  Therefore the Hubble sphere is at a constant proper distance and coincident with the event horizon.  This is also the late time asymptotic behaviour of eternally expanding FRW models with $\oll>0$ (see Fig.~\ref{fig:dist}, upper panel).}.  
Thus photons near the Hubble sphere that are receding slowly are overtaken by the more rapidly receding Hubble sphere\footnote{The myth that superluminally receding galaxies are beyond our view, may have propagated through some historical preconceptions.  Firstly, objects on our event horizon {\em do} have infinite redshift, tempting us to apply our SR knowledge that infinite redshift corresponds to a velocity of $c$.  Secondly, the once popular steady state theory predicts exponential expansion, for which the Hubble sphere and event horizon {\em are} coincident.}.


Our teardrop shaped past light cone in the top panel of Fig.~\ref{fig:dist} shows that any photons we now observe that were emitted in the first $\sim$ five billion years were emitted in regions that were receding superluminally, $v_{\rm rec}>c$.  Thus their total velocity was away from us.  Only when the Hubble sphere expands past these photons do they move into the region of subluminal recession and approach us. 
The most distant objects that we can see now were outside the Hubble sphere when their comoving coordinates intersected our past light cone. Thus, they were receding superluminally when they emitted the photons we see now. Since their worldlines have always been beyond the Hubble sphere these objects were, are, and always have been, receding from us faster than the speed of light. 

Evaluating Eq.~\ref{eq:vGR} for the observationally favoured $\omol=(0.3,0.7)$ universe shows that all galaxies beyond a redshift of $z=1.46$ are receding faster than the speed of light (Fig.~\ref{fig:vz}). 
Hundreds of galaxies with $z > 1.46$ have been observed.
The highest spectroscopic redshift observed in the Hubble deep field is $z=6.68$ (Chen \etal, 1999) 
and the Sloan digital sky survey has identified four galaxies at $z>6$ (Fan \etal, 2003). 
All of these galaxies have always been receding superluminally. The particle horizon, not the Hubble sphere, marks the size of our observable universe because we cannot have received light from, or sent light to, anything beyond the particle horizon\footnote{The current distance to our particle horizon and its velocity is difficult to say due to the unknown duration of inflation.  The particle horizon depicted in Fig.~\ref{fig:dist} assumes no inflation.}.  Our effective particle horizon is the cosmic microwave background (CMB), at redshift $z\sim 1100$, because we cannot see beyond the surface of last scattering.  Although the last scattering surface is not at any fixed comoving coordinate, the current recession velocity of the points from which the CMB was emitted is $3.2c$ (Fig.~\ref{fig:vz}).  At the time of emission their speed was $58.1c$, assuming $\omol=(0.3,0.7)$.  Thus we routinely observe objects that are receding faster than the speed of light and the Hubble sphere is not a horizon\footnote{Except in the special cases when the expansion is exponential, $R=R_0e^{Ht}$, such as the de Sitter universe ($\om=0,\oll>0$), during inflation or in the asymptotic limit of eternally expanding FRW universes.}.

\subsection{Ambiguity: The depiction of particle horizons on spacetime diagrams}
\label{sect:particlehorizon}

\begin{figure}
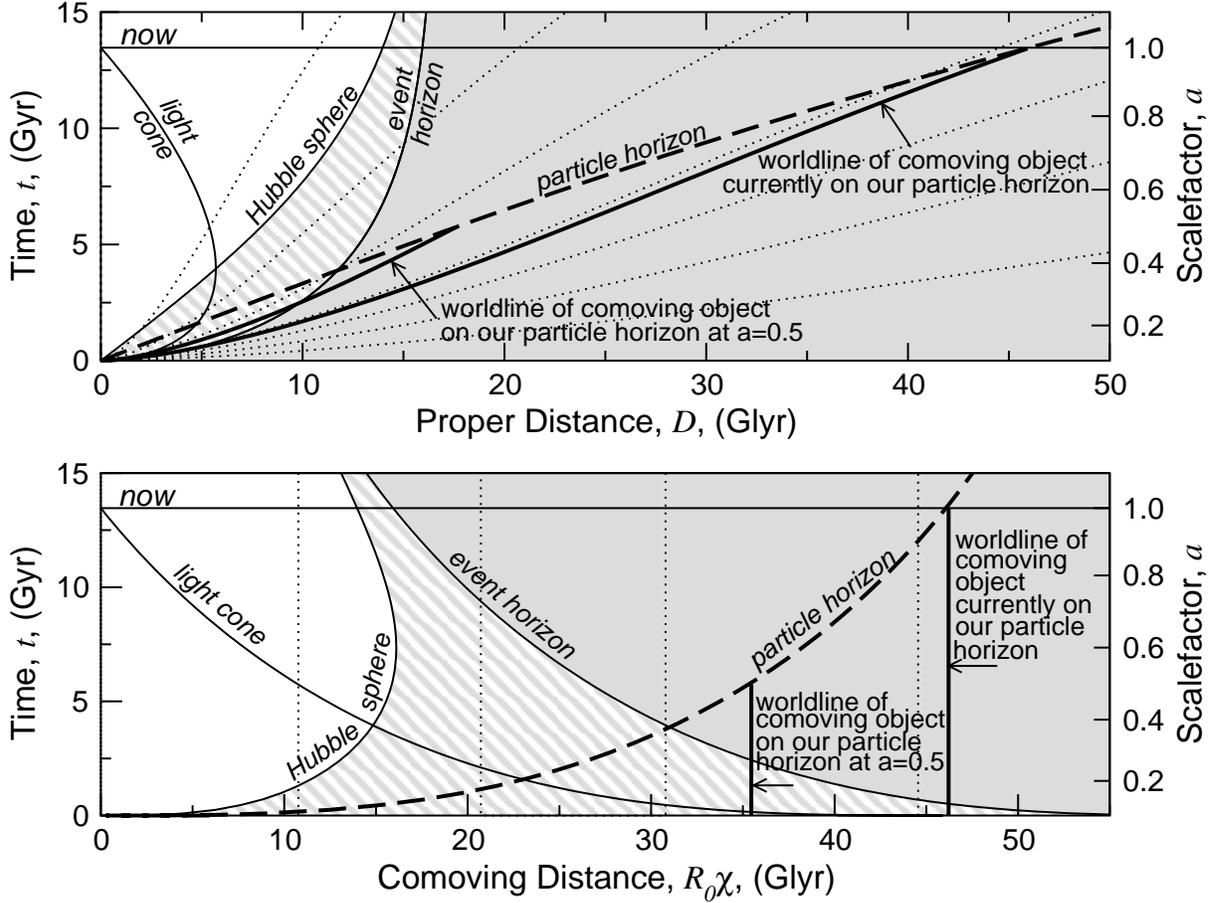
 \bctr
\psfig{file=fig3a.eps,width=160mm}\vspace{2mm}
\psfig{file=fig3b.eps,width=160mm}
\caption{The traditional depiction of the particle horizon on spacetime diagrams is the worldline of the object currently on our particle horizon (thick solid line).  All the information in this depiction is contained in a single point, the current distance to the particle horizon. An alternative way to plot the particle horizon is to plot the distance to the particle horizon as a function of time (thick dashed line and Fig.~\ref{fig:dist}).  This alleviates the need to draw a new worldline when we need to determine the particle horizon at another time (for example the worldline of the object on our particle horizon when the scalefactor $a=0.5$).}
\label{fig:ph}
\ectr \end{figure}

Here we identify an inconvenient feature of the most common depiction of the particle horizon on spacetime diagrams and provide a useful alternative (Fig.~\ref{fig:ph}).  The particle horizon at any particular time is a sphere around us whose radius equals the distance to the most distant object we can see. 
The particle horizon has traditionally been depicted as the {\em worldline} or comoving coordinate of the most distant particle that we have ever been able to see (Rindler, 1956; Ellis \& Rothman, 1993). 
The only information this gives is contained in a single point: the current distance of the particle horizon, and this indicates the current radius of the observable universe.   The rest of the worldline can be misleading as it does not represent a boundary between events we can see and events we cannot see, nor does it represent the distance to the particle horizon at different times.  An alternative way to represent the particle horizon is to plot the distance to the particle horizon as a function of time (Kiang, 1991). 
The particle horizon at any particular time defines a unique distance which appears as a single point on a spacetime diagram.  Connecting the points gives the distance to the particle horizon vs time.  It is this time dependent series of particle horizons that we plot in Fig.~\ref{fig:dist}.  (Rindler (1956) 
 calls this the boundary of our creation light cone -- a future light cone starting at the big bang.)  Drawn this way, one can read from the spacetime diagram the distance to the particle horizon at any time.  There is no need to draw another worldline.  

Specifically, what we plot as the particle horizon is $\chi_{\rm ph}(t)$ from Eq.~\ref{eq:chipht} rather than the traditional $\chi_{\rm ph}(t_0)$.  To calculate the distance to the particle horizon at an arbitrary time $t$ it is not sufficient to multiply $\chi_{\rm ph}(t_0)$ by $R(t)$ since the comoving distance to the particle horizon also changes with time.  

The particle horizon is sometimes distinguished from the event horizon by describing the particle horizon as a ``barrier in space'' and the event horizon as a ``barrier in spacetime''.  
This is not a useful distinction because both the particle horizon and event horizon are surfaces in spacetime -- they both form a sphere around us whose radius varies with time.   
When viewed in comoving coordinates the particle horizon and event horizon are mirror images of each other (symmetry about $z=10$ in the middle and lower panels of Fig.~\ref{fig:dist}). 
The traditional depiction of the particle horizon would appear as a straight vertical line in comoving coordinates, i.e.,~the comoving coordinate of the present day particle horizon (Fig.~\ref{fig:ph}, lower panel).

The proper distance to the particle horizon is {\em not} $D_{\rm PH}=ct_0$.  Rather, it is the proper distance to the most distant object we can observe, and is therefore related to how much the universe has expanded, i.e.~how far away the emitting object has become, since the beginning of time.  In general this is $\sim 3 c t_0$. The relationship between the particle horizon and light travel time arises because the comoving coordinate of the most distant object we can see {\em is} determined by the {\em comoving} distance light has travelled during the lifetime of the universe (Eq.~\ref{eq:chipht}).

\section{Observational evidence for the general relativistic interpretaion of cosmological redshifts}\label{sect:sr}
\label{sect:data}

\subsection{Duration-redshift relation for Type Ia Supernovae}

Many misconceptions arise from the idea that recession velocities are limited by SR to less than the speed of light so in Section~\ref{sect:mag-z} we present an analysis of supernovae observations yielding evidence against the SR interpretation of cosmological redshifts.
But first we would like to present an observational test that {\em can not} distinguish between special relativistic and general relativistic expansion of the Universe.  

General relativistic cosmology predicts that events occurring on a receding emitter will appear time dilated by a factor,
\beq \gamma_{\rm GR}(z) = 1+z. \eeq
A process that takes $\Delta t_0$ as measured by the emitter appears to take $\Delta t = \gamma_{\rm GR}\Delta t_0$ as measured by the observer when the light emitted by that process reaches them.
Wilson (1939) 
suggested measuring this cosmological time dilation to test whether the expansion of the Universe was the cause of cosmological redshifts. 
Type Ia supernovae (SNe Ia) lightcurves provide convenient standard clocks with which to test cosmological time dilation.   Recent evidence from supernovae includes Leibundgut \etal~(1996) 
who gave evidence for GR time dilation using a single high-$z$ 
supernova and Riess \etal~(1997) 
who showed  $1+z$ time dilation for a single SN Ia  at the 96.4\% confidence level using the time variation of spectral features.  Goldhaber \etal~(1997) 
show five data points of lightcurve width consistent with $1+z$ broadening and extend this analysis in Goldhaber \etal~(2001) 
to rule out any theory that predicts zero time dilation (for example ``tired light'' scenarios (see Wright, 2001)), 
at a confidence level of $18\sigma$. 
All of these tests show that $\gamma=(1+z)$ time dilation is preferred over models that predict {\em no} time dilation.   

We want to know whether the same observational test can show that GR time dilation is preferred over SR time dilation as the explanation for cosmological redshifts.  When we talk about SR expansion of the universe we are assuming that we have an inertial frame that extends to infinity (impossible in the GR picture) and that the expansion involves objects moving through this inertial frame.  The time dilation factor in SR is,
\bea \gamma_{\rm SR}(z) &=& (1-v_{\rm pec}^2/c^2)^{-1/2},\\
               &=&  \frac{1}{2}(1+z+\frac{1}{1+z})\approx 1 + z^2/2.\label{eq:gammaSRz}\eea
This time dilation factor relates the proper time in the moving emitter's inertial frame ($\Delta t_0$) to the proper time in the observer's inertial frame ($\Delta t_1$).  To measure this time dilation the observer has to set up a set of synchronized clocks (each at rest in the observer's inertial frame) and take readings of the emitter's proper time as the emitter moves past each synchronized clock.  The readings show that the emitter's clock is time dilated such that $\Delta t_1 = \gamma_{\rm SR} \Delta t_0$. 

We do not have this set of synchronized clocks at our disposal when we measure time dilation of supernovae and therefore Eq.~\ref{eq:gammaSRz} is not the time dilation we observe.  In an earlier version of this paper we mistakenly attempted to use this equation to show SR disagreed with observational results.  This could be classed as an example of an ``expanding confusion''.
For the observed time dilation of supernovae we have to take into account an extra time dilation factor that occurs because the distance to the emitter (and thus the distance light has to propagate to reach us) is increasing.   In the time $\Delta t_1$ the emitter moves a distance $v\Delta t_1$ away from us.  The total proper time we observe ($\Delta t$) is $\Delta t_1$ plus an extra factor describing how long light takes to traverse this extra distance ($v\Delta t_1/c$),
\beq \Delta t = \Delta t_1 (1+v/c). \eeq
The relationship between proper time at the emitter and proper time at the observer is thus,
\bea \Delta t &=& \Delta t_0 \gamma (1+v/c), \\
              &=& \Delta t_0 \sqrt{\frac{1+v/c}{1-v/c}},\\
              &=& \Delta t_0 (1+z).\eea
This is identical to the GR time dilation equation.  Therefore using time dilation to distinguish between GR and SR expansion is impossible.

Leibundgut \etal~(1996), 
Riess \etal~(1997) 
and Goldhaber \etal~(1997, 2001) 
do provide excellent evidence that expansion is a good explanation for cosmological redshifts.  What they can not show is that GR is a better description of the expansion than SR.  Nevertheless, other observational tests provide strong evidence against the SR interpretation of cosmological redshifts, and we demonstrate one such test in the next section.

\begin{figure}[h]\bctr
\psfig{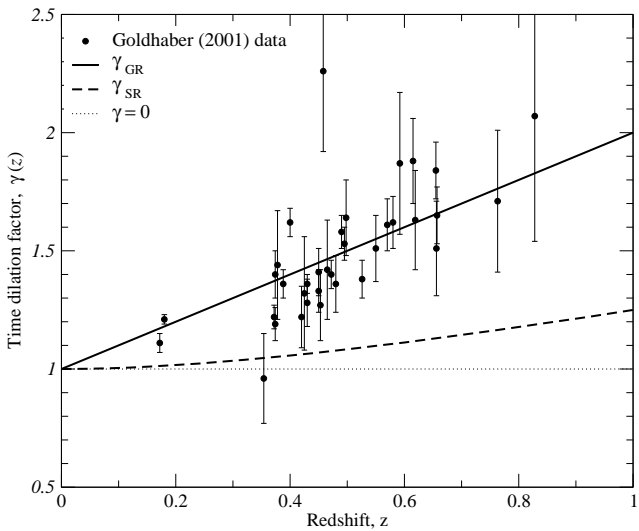}
\caption{\fns Supernovae time dilation factor vs redshift.  The solid line is the time dilation factor predicted by both general relativity and special relativity.  The thick dashed line is the special relativistic time dilation factor that a set of synchronized clocks spread throughout our inertial frame would observe, without taking into account the changing distance light has to travel to reach us.  Once the change in the emitter's distance is taken into account SR predicts the same time dilation effect as GR, $\gamma = (1+z)$.  The thin dotted line represents any theory that predicts no time dilation (e.g. tired light). The 35 data points are from Goldhaber \etal~(2001).    
They rule out no time dilation at a confidence level of $18\sigma$.  
}
\label{fig:Goldhaber}\ectr
\end{figure}

\begin{figure} \bctr
\psfig{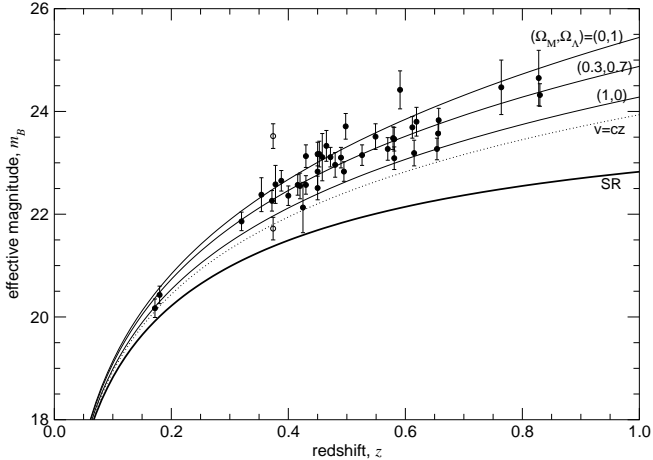}
\caption{Magnitude-redshift relation for several models with data taken from Perlmutter et al. 1999 [Fig.~2(a)].  The SR prediction has been added (as described in text), as has the prediction assuming a linear $v=cz$ relationship.  The interpretation of the cosmological redshift as an SR Doppler effect is  ruled out at more than $23\sigma$ compared with the $\Lambda$CDM concordance model. 
The linear $v=cz$ model is a better approximation than SR, but is still ruled out at $12\sigma$.  
}
\label{fig:mag-z}
\ectr \end{figure}
\vspace{1cm}
\subsection{Magnitude-redshift relationship for SNe Ia}\label{sect:mag-z}
Another observational confirmation of the GR interpretation that {\em is} able to rule out the SR interpretation is the curve in the magnitude-redshift relation.  SNe Ia are being used as standard candles to fit the magnitude-redshift relation out to redshifts close to one (Riess \etal, 1998; Perlmutter \etal, 1999). 
Recent measurements are accurate enough to put restrictions on the cosmological parameters $\omol$.  We perform a simple analysis of the supernovae magnitude-redshift data to show that it also strongly excludes the SR interpretation of cosmological redshifts (Fig.~\ref{fig:mag-z}).

Figure~\ref{fig:mag-z} shows the theoretical curves for several GR models accompanied by the observed SNe Ia data from Perlmutter \etal~(1999) 
 [their Fig.~2(a)].  The conversion between luminosity distance, $D_L$ (Eq.~\ref{eq:luminosity}), and effective magnitude in the B-band given in Perlmutter \etal~(1999), 
is $m_{\rm B}(z)=5\log H_0 D_L + M_{\rm B}$ where $M_{\rm B}$ is the absolute magnitude in the $B$-band at the maximum of the light curve. They marginalize over $M_{\rm B}$ in their statistical analyses.  We have taken $M_{\rm B} =-3.45$ which closely approximates their plotted curves.  

We superpose the curve deduced by interpreting Hubble's law special relativistically.  One of the strongest arguments against using SR to interpret cosmological redshifts is the difficulty in interpreting observational features such as magnitude.    We calculate $D(z)$ special relativistically by assuming the velocity in $v=HD$ is related to redshift via Eq.~\ref{eq:vSR}, so, 
\beq D(z)=\frac{c}{H}\frac{(1+z)^2-1}{(1+z)^2+1}.\eeq  
Since all the redshifting happens at emission in the SR scenario, $v$ should be calculated at the time of emission.  However, since SR does not provide a technique for incorporating acceleration into our calculations for the expansion of the Universe,  the best we can do is assume that the recession velocity, and thus Hubble's constant, are approximately the same at the time of emission as they are now\footnote{There are several complications that this analysis does not address.  (1) SR could be manipulated to give an evolving Hubble's constant and (2) SR could be manipulated to give a non-trivial relationship between luminosity distance, $D_L$, and proper distance, $D$.  However, it is not clear how one would justify these ad hoc corrections.}.  We then convert $D(z)$ to $D_L(z)$ using Eq.~\ref{eq:luminosity}, so $D_L(z)=D(z)(1+z)$.  This version of luminosity distance has been used to calculate $m(z)$ for the SR case in Fig.~\ref{fig:mag-z}. 

SR fails this observational test dramatically being $23\sigma$ from the general relativistic $\Lambda$CDM model $\omol=(0.3,0.7)$.  
We also include the result of assuming $v=cz$.  Equating this to Hubble's law gives, $D_L(z)=cz(1+z)/H$.  For this observational test the linear prediction is closer to the GR prediction (and to the data) than SR is.  Nevertheless the linear result lies $12\sigma$ from the $\Lambda$CDM concordance result.




\subsection{Future tests}\label{sect:futuretests}
\begin{figure}[!t] \bctr
\psfig{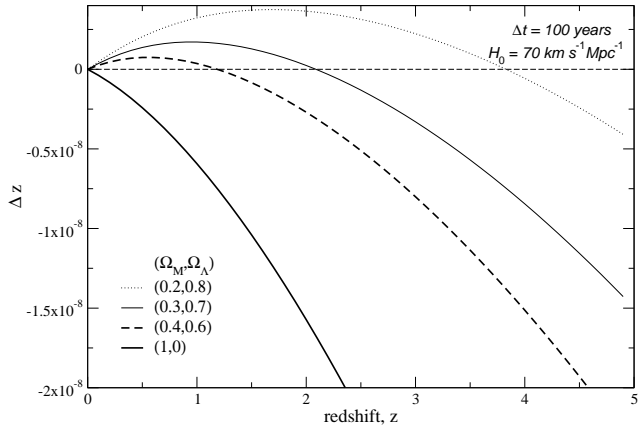}
\caption{The change in the redshift  of a comoving object as predicted by FRW cosmology for various 
cosmological models.  The horizontal axis represents the initial redshifts.  The timescale taken for the change is 
100 years.  The changes predicted are too small for current instrumentation to detect.
}
\label{fig:dz-z}
\ectr \end{figure}

Current instrumentation is not accurate enough to perform some other observational tests of GR.  For example Sandage (1962) 
showed that the evolution in redshift of distant galaxies due to the acceleration or deceleration of the universe is a direct way to measure the cosmological parameters.  The change in redshift over a time interval $t_0$ is given by,
\beq \frac{dz}{dt_0} = H_0(1+z) - H_{\rm em},\label{eq:dz}\eeq 
where $H_{\rm em}=\dot{R}_{\rm em}/R_{\rm em}$ is Hubble's constant at the time of emission.  Unfortunately the magnitude of the redshift variation is small over human timescales.  Ebert \& Tr\"umper (1975), Lake (1981), Loeb (1998) 
and references therein each reconfirmed that the technology of the day did not yet provide precise enough redshifts to make such an observation viable.  Figure~\ref{fig:dz-z} shows the expected change in redshift due to cosmological acceleration or deceleration is only $\Delta z \sim 10^{-8}$ over 100 years.   Current Keck/HIRES spectra with iodine cell reference wavelengths can measure quasar absorption line redshifts to an accuracy of $\Delta z \sim 10^{-5}$ (Outram \etal, 1999). 
Thus, this observational test must wait for future technology.

\section{Discussion}\label{sect:discussion}

Recession velocities of individual galaxies are of limited use in {\em observational} cosmology because they are not directly observable.  For this reason some of the physics community considers recession velocities meaningless and would like to see the issue swept under the rug~\incite{24--25}. 
They argue that we should refrain from interpreting observations in terms of velocity or distance, and stick to the observable, redshift.  This avoids any complications with superluminal recession and avoids any confusion between the variety of observationally-motivated definitions of distance commonly used in cosmology (e.g. Eqs.~\ref{eq:luminosity} and~\ref{eq:angsize}).  

However, redshift is not the only observable that indicates distance and velocity.  The host of low redshift distance measures and the multitude of available evidence for the Big Bang model all suggest that higher redshift galaxies are more distant from us and receding faster than lower redshift galaxies.  Moreover, we cannot currently sweep distance and velocity under the rug if we want to explain the cosmological redshift itself.  Expansion has no meaning without well-defined concepts of velocity and distance.  If recession velocity were meaningless we could not refer to an ``expanding universe'' and would have to restrict ourselves to some operational description such as ``fainter objects have larger redshifts''.  However, within general relativity the relationship between cosmological redshift and recession velocity is straightforward.  Observations of SNe Ia 
apparent magnitudes provide independent evidence that the cosmological redshifts are due to the general relativistic expansion of the universe. 
Understanding distance and velocity is therefore fundamental to the understanding of our Universe. 

When distances are large enough that light has taken a substantial fraction of the age of the Universe to reach us there are more observationally convenient distance measures than proper distance, such as luminosity distance and angular-size distance.  The most convenient distance measure depends on the method of observation.  Nevertheless, these distance measures can be converted between each other, and so collectively define a unique concept\footnote{\vspace{-5mm}\bea \mbox{Proper Distance} & D  = &R \chi\\
   \mbox{Luminosity Distance}& D_L= &R S_k(\chi)(1+z)\label{eq:luminosity}\\
   \mbox{Angular-size Distance}& D_{\theta}= &R S_k(\chi)(1+z)^{-1}\label{eq:angsize}\eea }. In this paper we have taken proper distance to be the fundamental {\em radial} distance measure.    Proper distance is the spatial geodesic measured along a hypersurface of constant cosmic time (as defined in the Robertson-Walker metric).  It is the distance measured along a line of sight by a series of infinitesimal comoving rulers at a particular time, $t$.  Both luminosity and angular-size distances are calculated from observables involving distance perpendicular to the line of sight and so contain the angular coefficient $S_k(\chi)$.  They parametrize radial distances but are not geodesic distances along the three dimensional spatial manifold\footnote{Note also that the standard definition of angular-size distance is purported to be the physical size of an object, divided by the angle it subtends on the sky.  The physical size used in this equation is not actually a length along a spatial geodesic, but rather along a line of constant $\chi$ (Liske, 2000). 
The correction is negligible for the small angles usually measured in astronomy.}.  They are therefore not relevant for the calculation of recession velocity\footnote{Murdoch, H.~S. 1977, ``[McVittie] regards as equally valid other definitions of distance such as luminosity distance and distance by apparent size.  But while these are extremely useful concepts, they are really only definitions of observational convenience which extrapolate results such as the inverse square law beyond their range of validity in an expanding universe.''}. 
Nevertheless, if they were used, our results would be similar.  Only angular-size distance can avoid superluminal velocities because $D_\theta=0$ for both $z=0$ and $z\rightarrow \infty$ (Murdoch, 1977). 
Even then the rate of change of angular-size distance does not approach $c$ for $z\rightarrow \infty$. 

Throughout this paper we have used proper time, $t$, as the temporal measure.  This is the time that appears in the RW metric and the Friedmann equations. This is a convenient time measure because it is the proper time of comoving 
observers.  Moreover, the homogeneity of the universe is dependent on this choice of time coordinate --- if any other time coordinate were chosen (that is not a trivial multiple of $t$) the density of the universe would be distance dependent.  
Time can be defined differently, for example to make the SR Doppler shift formula  (Eq.~\ref{eq:vSR}) correctly calculate recession velocities from observed redshifts (Page, 1993). 
However, to do this we would have to sacrifice the homogeneity of the universe and the synchronous proper time of comoving objects (Davis \& Lineweaver, 2003). 

\section{Conclusion}
We have clarified some common misconceptions surrounding the expansion of the universe, and shown with numerous references how misleading statements manifest themselves in the literature.  
Superluminal recession is a feature of all expanding cosmological models that are homogeneous and isotropic and therefore obey Hubble's law.  This does not contradict special relativity because the superluminal motion does not occur in any observer's inertial frame.  All observers measure light locally to be travelling at $c$ and nothing ever overtakes a photon.  Inflation is often called ``superluminal recession'' but even during inflation objects with $D<c/H$ recede subluminally while objects with $D>c/H$ recede superluminally.  Precisely the same relationship holds for non-inflationary expansion.  We showed that the Hubble sphere is not a horizon --- we routinely observe galaxies that have, and always have had, superluminal recession velocities.  All galaxies at redshifts greater than $z\sim 1.46$ today are receding superluminally in the $\Lambda$CDM concordance model.  We have also provided a more informative way of depicting the particle horizon on a spacetime diagram than the traditional worldline method.  
An abundance of observational evidence supports the general relativistic big bang model of the universe.  The duration of supernovae light curves shows that models predicting no expansion are in conflict with observation.
Using magnitude-redshift data from supernovae 
we were able to rule out the SR interpretation of cosmological redshifts at the $\sim23\sigma$ level.   
Together these observations provide strong evidence that 
the general relativistic interpretation of the cosmological redshifts is preferred over special relativistic and tired light interpretations.  The general relativistic description of the expansion of the universe agrees with observations, and does not need any modifications for $\vrec > c$. 


\appendix

\section{Standard general relativistic definitions of expansion and horizons}\label{sect:math}

The metric for an homogeneous, isotropic universe is the Robertson-Walker (RW) metric,
\beq  
ds^2 = -c^2dt^2 + R(t)^2[d\chi^2+S_k^2(\chi)d\psi^2], 
\label{eq:frwmetric}
\eeq
where $c$ is the speed of light, $dt$ is the time separation, $d\chi$ is the comoving coordinate separation and $d\psi^2=d\theta^2+\sin^2\theta d\phi^2$, where $\theta$ and $\phi$ are the polar and azimuthal angles in spherical coordinates. The scalefactor, $R$, has dimensions of distance.   The function $S_k(\chi)=\sin\chi$, $\chi$ or $\sinh\chi$ for closed ($k=+1$), flat ($k=0$) or open ($k=-1$) universes respectively (Peacock, 1999, p.~69). 
The proper distance $D$, at time $t$, in an expanding universe, between an observer at the origin and 
a distant galaxy is defined to be along a surface of constant time ($dt=0$).  
We are interested in the radial distance so $d\psi=0$. The RW metric then reduces to $ds = R d\chi$ which, upon 
integration yields,
\beq {\rm Proper} \; {\rm distance,}\quad\quad\quad \;D(t) = R(t)\chi.  \label{eq:Hubble}\eeq
Differentiating this yields the theoretical form of Hubble's law (Harrison, 1993), 
\bea{\rm Recession}\; {\rm velocity,}\,\;\;v_{\rm rec}(t,z) &=& \dot{R}(t)\chi(z), \label{eq:vchi}\\
&=& H(t) D(t),  \label{eq:hubbleslaw}
\eea
where $v_{\rm rec}=\dot{D}$ (for $\dot{\chi}=0$) and $\chi(z)$ is the fixed comoving coordinate associated with a galaxy observed today at redshift $z$.  Note that the redshift of an object at this fixed comoving coordinate changes with time\footnote{In addition, objects that have a peculiar velocity also move through comoving coordinates.   Therefore more generally Eq.~\ref{eq:vchi} above should be written with $\chi$ explicitly time dependent, $v_{\rm rec}(t,z)=\dot{R}(t)\chi(z,t)$.} (Eq.~\ref{eq:dz}).
A distant galaxy will have a particular recession velocity when it emits the photon at $t_{\rm em}$ and a different recession velocity when we observe the photon at $t_0$. Eq.~\ref{eq:hubbleslaw} evaluated at $t_0$ gives the recession velocities plotted in Fig.~\ref{fig:vz}.

The recession velocity of a comoving galaxy is a time dependent quantity because the expansion rate of the 
universe $\dot{R}(t)$ changes with time.  The current recession 
velocity of a galaxy is given by $v_{\rm rec}= \dot{R}_0\chi(z)$.  
On the spacetime diagram of Fig.~\ref{fig:dist} this is the velocity taken at points along the line of constant 
time marked ``now''.
The recession velocity of an emitter at the time it emitted the light we observe is the velocity at points along our the past light cone\footnote{The recession velocity at the time of emission is $v_{\rm rec}(t_{\rm em}) = R(t_{\rm em})\chi(z)$ where $R(t_{\rm em})=R(t)$ as defined in Eq.~\ref{eq:z}.}.  However, we can also compute the recession 
velocity a comoving object has at {\em any time} during the history of the universe, having initially calculated 
its comoving coordinate from its present day redshift.

Allowing $\chi$ to vary when differentiating Eq.~\ref{eq:Hubble} with respect to time gives two distinct velocity terms (Landsberg \& Evans, 1977; Silverman, 1986; Peacock, 1999; Davis \etal, 2003), 
\bea \dot{D} &=& \dot{R}\chi + R\dot{\chi}, \\
     v_{\rm tot} &=& v_{\rm rec} + v_{\rm pec}.\label{eq:vtot}\eea
This explains the changing slope of our past light cone in the upper panel of Fig.~\ref{fig:dist}.  The peculiar velocity of light is always $c$ (Eq.~\ref{eq:dchi}) so the total velocity of light whose peculiar velocity is towards us is $v_{\rm tot}=v_{\rm rec}-c$ 
which is always positive (away from us) when $v_{\rm rec}>c$.  Nevertheless we can eventually receive photons that initially were receding from us because the Hubble sphere expands and overtakes the receding photons so the photons find themselves in a region with $v_{\rm rec}<c$ (Section~\ref{sect:notobserve}).

Photons travel along null geodesics, $ds=0$.  To obtain the comoving distance, $\chi$, between an observer at the origin 
and a galaxy observed to have a redshift $z(t)$, set $ds = 0$ (to measure along the path of a photon) and $d\psi = 0$ 
(to measure radial distances) in the RW metric yielding, 
\beq c\;dt = R(t) d\chi.  \label{eq:dchi} \eeq
This expression confirms our previous statement that the {\em peculiar} velocity of a photon, $R\dot{\chi}$, is $c$.  
Since the velocity of light through comoving coordinates is not constant ($\dot{\chi}=c/R$), to 
calculate comoving distance we cannot simply multiply the speed of light through comoving space by time.  We have 
to integrate over this changing comoving speed of light for the duration of propagation.  Thus, the comoving coordinate of a 
comoving object that emitted the light we now see at time $t$ is attained by integrating Eq.~\ref{eq:dchi},
\beq 
{\rm Past} \;{\rm Light} \;{\rm  Cone,}\: \:
\chi_{\rm lc}(t_{\rm em}) = c\int_{t_{\rm em}}^{t_0}\frac{dt'}{R(t')}.\label{eq:lightconet}
\eeq
We can parametrize time using redshift and thus recast Eq.~\ref{eq:lightconet} in terms of observables.
The cosmological redshift of an object is given by the ratio of the scalefactor at the time of observation, $R(t_0)=R_0$, to the scalefactor at the time of emission, $R(t)$,
\beq 
{\rm Redshift,}\;\;\;    1+z = \frac{R_{0}}{R(t)}. 
\label{eq:z}\eeq
Differentiating  Eq.~\ref{eq:z} with respect to $t$ gives $dt/R(t)=-dz/R_{\rm 0}H(z)$ where redshift is used instead of time to parametrize Hubble's constant.  $H(z)$ is Hubble's constant at the time an object with redshift, $z$, emitted the light we now see.  Thus, for the limits of the integral in Eq.~\ref{eq:lightconet}, the time of emission becomes $z=0$ while the time of observation becomes the observed redshift, $z$.
The comoving coordinate of an object in terms of observables is therefore,
\beq
\chi(z) = \frac{c}{R_0}\;\int_{o}^{z} \frac{dz^{\prime}}{H(z^{\prime})}.  \label{eq:chiz}
\eeq
Thus, there is a direct one to one relationship between observed redshift and comoving coordinate.   Notice that in 
contrast to special relativity, the redshift does not indicate the velocity, it indicates the distance\footnote{\footnotesize Distance is proportional to recession velocity at any particular time, but a particular 
redshift measured at different times will correspond to different recession velocities.}.  That is, the redshift tells us not the velocity of the emitter, but where the emitter sits (at rest locally) in the coordinates of the universe.  The recession velocity is obtained by inserting Eq.~\ref{eq:chiz} into Eq.~\ref{eq:vchi} yielding Eq~\ref{eq:vGR}. 

The Friedmann Equation gives the time dependence of Hubble's constant,
\beq H(z)=H_0 \: (1+z)\left[1+\om z+\oll\left(\frac{1}{(1+z)^2}-1\right)\right]^{1/2}.\label{eq:H} \eeq    
Expressing Hubble's constant this way is useful because it is in terms of observables.  However, it restricts our calculations to objects with redshift $z<\infty$.  That is, objects we can currently see.  There is no reason to assume the universe ceases beyond our current particle horizon.  Expressing Hubble's constant as $H(t)=\dot{R}(t)/R(t)$ allows us to extend the analysis to a time of observation, $t\rightarrow\infty$, which is beyond what we can currently observe.  Friedmann's equation is then (using the scalefactor normalized to one at the present day $a(t)=R(t)/R_0$), 
\beq 
\dot{R}(t) = R_0H_0\left[1+\om\left(\frac{1}{a}-1\right)+\oll(a^2-1)\right]^{1/2},\label{eq:fried}
\eeq
which we use with the identity $dt/R(t) = dR/(\dot{R}R)$ to evaluate Eqs.~\ref{eq:lightconet},~\ref{eq:chipht} and~\ref{eq:eventhorizont}.

Altering the limits on the integral in Eq.~\ref{eq:lightconet} gives the horizons we have plotted on the spacetime diagrams.
The time dependent particle horizon we plot in Fig.~\ref{fig:dist} uses $D_{\rm ph}=R(t)\chi_{\rm ph}(t)$ with,
\beq 
{\rm Particle}\:\:{\rm Horizon,}\: \:\chi_{\rm ph}(t)= c\int_0^{t}\frac{dt^\prime}{R(t^\prime)}. 
\label{eq:chipht}
\eeq
The traditional depiction of the particle horizon as a worldline uses $D_{\rm ph}=R(t)\chi_{\rm ph}(t_0)$.  
The comoving distance to the event horizon is given by,
\beq 
{\rm Event}\:\:{\rm  Horizon,} \;\;\:\chi_{\rm eh}(t)= c\int_{t}^{t_{\rm end}}\frac{dt^{\prime}}{R(t^{\prime})},
\label{eq:eventhorizont}
\eeq
where $t_{\rm end}=\infty$ in eternally expanding models or the time of the big crunch in recollapsing models.  

A conformal time interval, $d\tau$, is defined as a proper time interval $dt$ divided by the scalefactor,
\beq 
{\rm Conformal}\;\;{\rm time,}\;\;\;   d\tau = dt/R(t). \label{eq:conformalt} \eeq




\vspace{1cm}
\section{Examples of misconceptions or easily misinterpreted statements in the literature}\label{sect:quotes}
In text books and works of popular science it is often standard practice to simplify arguments for the reader.  Some of the quotes below fall into this category.  We include them here to point out the difficulty encountered by someone starting in this field and trying to decipher what is really meant by `the expansion of the Universe'.  \vspace{3mm}

\noindent
{\footnotesize 
\xcite{1}{Feynman, R.~P. 1995, {\em Feynman Lectures on Gravitation (1962/63)}, (Reading, Mass.: Addison-Wesley) p.~181, ``It makes no sense to worry about the possibility of galaxies receding from us faster than light, whatever that means, since they would never be observable by hypothesis.''}
\xcite{2}{Rindler, W. 1956,  \mnras, {\bf 6}, 662-667, {\em Visual Horizons in World-Models}, Rindler acknowledged that faster than $c$ expansion is implicit in the mathematics, but expresses discomfort with the concept: ``\dots certain physical difficulties seem to be inherent in models possessing a particle-horizon: if the model postulates point-creation we have material particles initially separating at speeds exceeding those of photons.''}
\xcite{3}{McVittie, G.~C. 1974, \qjras, {\bf 15}, 246-263, {\em Distances and large redshifts}, Sect.~4, ``These fallacious arguments would apparently show that many quasars had `velocities of recession' greater than that of light, which contradicts one of the basic postulates of relativity theory.''}
\xcite{4}{Weinberg, S. 1977, {\em The First Three Minutes}, (New York: Bantum Books), p.~27, ``The conclusion generally drawn from this half century of observation is that the galaxies are receding from us, with speeds proportional to the distance (at least for speeds not too close to that of light).'', see also p.~12 and p.~25.  Weinberg makes a similar statement in his 1972 text {\em Gravitation and Cosmology} (New York: Wiley), p.~417, ``a {\em relatively close} galaxy will move away from or toward the Milky Way, with a radial velocity [$v_{\rm rec}=\dot{R}(t_0)\chi$].'' (emphasis ours).  Shortly thereafter he adds a caution about SR and distant sources: ``it is neither useful nor strictly correct to interpret the frequency shifts of light from very distant sources in terms of a special-relativistic D\"oppler shift alone. [The reader should be warned though, that astronomers conventionally report even large frequency shifts in terms of a recessional velocity, a ``red shift'' of $v$ km/sec meaning that $z=v/(3\times 10^5)$.]''}
\xcite{5}{Field, G. 1981, {\em This Special Galaxy}, in Section II of {\em Fire of life, the book of the Sun}, (Washington, DC: Smithsonian Books) ``The entire universe is only a fraction of a kilometer across [after the first millionth of a second], but it expands at huge speeds --- matter quite close to us being propelled at almost the speed of light.''}
\xcite{6}{Schutz, B.~F. 1985, {\em A first course in General Relativity}, (\cupadr: \cup) p.~320, ``[v=HD] cannot be exact since, for $D>1.2\times10^{26}m=4000$ Mpc, the velocity exceeds the velocity of light!  These objections are right on both counts.  Our discussion was a {\em local} one (applicable for recession velocity $<<1$) and took the point of view of a particular observer, ourselves.  Fortunately, the cosmological expansion is slow...''}
\xcite{7}{Peebles, P.~J.~E., Schramm, D.~N., Turner, E.~L. and Kron, R.~G. 1991, Nature {\bf 352}, 769, {\em The case for the relativistic hot Big Bang cosmology}, ``There are relativistic corrections [to Hubble's Law, $v=H_0D$,] when $v$ is comparable to the velocity of light $c$.''  However, Peebles, in his 1993 text {\em Principles of Physical Cosmology}, (Princeton: Princeton University Press), p.~98, explains: ``Since equation [$D=R\chi$] for the proper distance [$D$] between two objects is valid whatever the coordinate separation, we can apply it to a pair of galaxies with separation greater than the Hubble length... Here the rate of change of the proper separation, [$\dot{D}=HD$], is greater than the velocity of light.  This is not a violation of special relativity;''  Moreover, in the next paragraph Peebles makes it clear that, dependent upon the cosmological parameters, we can actually observe objects receding faster than the speed of light.}
\xcite{8}{Peacock, J.~A. 1999, {\em Cosmological Physics}, (\cupadr: \cup) p.~6, ``\dots objects at a vector distance {\bf r} appear to recede from us at a velocity ${\bf v}=H_0{\bf r}$, where $H_0$ is known as Hubble's constant (and is not constant at all as will become apparent later.)  This law is only strictly valid at small distances, of course, but it does tell us that objects with $r\simeq c/H_0$ recede at a speed approaching that of light.  This is why it seems reasonable to use this as an upper cutoff in the radial part of the above integral.''  However, Peacock makes it very clear that cosmological redshifts are not due to the special relativistic Doppler shift, p72, ``it is common but misleading to convert a large redshift to a recession velocity using the special-relativistic formula $1+z = [(1+v/c)/(1-v/c)]^{1/2}$.  Any such temptation should be avoided'' }
%
\xcite{9}{Davies, P.~C.~W. 1978, {\em The Runaway Universe} (London: J. M. Dent \& Sons Ltd) p.~26, ``\dots galaxies several billion light years away seem to be increasing their separation from us at nearly the speed of light.  As we probe still farther into space the redshift grows without limit, and the galaxies seem to fade out and become black.  When the speed of recession reaches the speed of light we cannot see them at all, for no light can reach us from the region beyond which the expansion is faster than light itself.  This limit is called our horizon in space, and separates the regions of the universe of which we can know from the regions beyond about which no information is available, however powerful the instruments we use.''}
\xcite{10}{Berry, M. 1989, {\em Principles of Cosmology and Gravitation}, (Bristol, U.K.: IOP Publishing) p.~22 ``\dots if we assume that Euclidean geometry may be employed, \dots galaxies at a distance  $D_{max} = c/H \sim 2 \times 10^{10}$ light years $\sim 6\times 10^9$ pc are receding as fast as light.  Light from more distant galaxies can never reach us, so that $D_{max}$ marks the limit of the observable universe; it corresponds to a sort of {\em horizon}.''}
\xcite{11}{Raine, D.~J. 1981, {\em The Isotropic Universe}, (Bristol: Adam Hilber Ltd) p.~87, ``One might suspect special relativistic effects to be important since some quasars are observed to exhibit redshifts, $z$, in excess of unity.  This is incompatible with a Newtonian interpretation of the Doppler effect, since one would obtain velocities $v=cz$ in excess of that of light.  The special relativistic Doppler formula $1+z=(c+v)/(c-v)^{1/2}$ always leads to sub-luminal velocities for objects with arbitrarily large redshifts, and is at least consistent.  In fact we shall find that the strict special relativistic interpretation is also inadequate.  Nevertheless, at the theoretical edge of the visible Universe we expect at least in principle to see bodies apparently receding with the speed of light.''}
\xcite{12}{Liddle, A.~R. 1988, {\em An introduction to Modern Cosmology}, (Sussex: John Wiley \& Sons Ltd) p.~23, Sect 3.3, ``\dots ants which are far apart on the balloon could easily be moving apart at faster than two centimetres per second if the balloon is blown up fast enough.  But if they are, they will never get to tell each other about it, because the balloon is pulling them apart faster than they can move together, even at full speed.''}
\xcite{13}{Krauss, L.~M. and Starkman, G.~D. 1999, \apj, 531(1), 22--30, {\em Life, the universe and nothing:  Life and death in an ever-expanding universe}, ``Equating this recession velocity to the speed of light $c$, one finds the physical distance to the so-called de Sitter horizon... This horizon, is a sphere enclosing a region, outside of which no new information can reach the observer at the center''.  This would be true if only applied to empty universes with a cosmological constant - de Sitter universes.  However this is not its usage: ``the universe became $\Lambda$-dominated at about 1/2 its present age.  The `in principle' observable region of the Universe has been shrinking ever since.  ... Objects more distant than the de Sitter horizon [Hubble Sphere] now will forever remain unobservable.''}
\xcite{14}{Harrison, E.~R. 1991, \apj, 383, 60--65, {\em Hubble spheres and particle horizons}, ``All accelerating universes, including universes having only a limited period of acceleration, have the property that galaxies at distances $L<L_H$ are later at $L>L_H$, and their subluminal recession in the course of time becomes superluminal.  Light emitted outside the Hubble sphere and traveling through space toward the observer recedes and can never enter the Hubble sphere and approach the observer.  Clearly, there are events that can never be observed, and such universes have event horizons.'' The misleading part of this quote is subtle -- there will be an event horizon in such universes (accelerating universes), but it needn't coincide with the Hubble sphere.  Unless the universe is accelerating so quickly that the Hubble sphere does not expand (exponential expansion) we will still observe things from beyond the Hubble sphere, even though there is an event horizon (see Fig.~\ref{fig:dist}).}
\xcite{15}{Harwit, M. 1998, {\em Astrophysical Concepts}, 3rd Ed., (New York: Springer-Verlag) p.~467, ``Statement (i) [In a model without an event horizon, a fundamental observer can sooner or later observe any event.] depends on the inability of particles to recede at a speed greater than light when no event horizon exists.''}
\xcite{16}{Hubble, E. and Humason, M.~L. 1931, \apj, 74, 443--480, {\em The Velocity-Distance relation among Extra-Galactic Nebulae}, pp.~71--72, ``If an actual velocity of recession is involved, an additional increment, equal to that given above, must be included in order to account for the difference in the rates at which the quanta leave the source and reach the observer$^1$.'', Footnote 1: ``The factor is $\sqrt{\frac{1+v/c}{1-v/c}}$ which closely approximates $1+d\lambda/\lambda$ for red-shifts as large as have been observed.  A third effect due to curvature, negligible for distances observable at present, is discussed by R. C. Tolman...''}
\xcite{17}{Lightman, A.~P., Press, W.~H., Price, R.~H. and Teukolsky, S.~A. 1975, {\em Problem book in relativity and gravitation}, (\pupadr: \pup) Prob.~19.7}
\xcite{18}{Halliday, D., Resnick, R. and Walker, J. 1974, {\em Fundamentals of Physics}, (USA: John Wiley \& Sons) 4th Ed., Question 34E, ``Some of the familiar hydrogen lines appear in the spectrum of quasar 3C9, but they are shifted so far toward the red that their wavelengths are observed to be three times as large as those observed for hydrogen atoms at rest in the laboratory.  (a) Show that the classical Doppler equation gives a relative velocity of recession greater than $c$ for this situation.  (b) Assuming that the relative motion of 3C9 and the Earth is due entirely to recession, find the recession speed that is predicted by the relativistic Doppler equation.'' See also Questions 28E, 29E and 33E.}
\xcite{19}{Seeds, M.~A. 1985, {\em Horizons - Exploring the Universe}, (Belmont, California: Wadsworth Publishing) pp.~386--387, ``If we use the classical Doppler formula, a red shift greater than 1 implies a velocity greater than the speed of light.  However, as explained in Box 17-1, very large red shifts require that we use the relativistic Doppler formula.... {\em Example:} Suppose a quasar has a red shift of 2.  What is its velocity?  {\em Solution:} [uses special relativity]''}
\xcite{20}{Kaufmann, W.~J. and Freedman, R.~A. 1988, {\em Universe}, (New York: W.~H.~Freeman \& Co.) Box 27-1, p.~675, ``\dots quasar PKS2000-330 has a redshift of $z=3.78$.  Using this value and applying the full, relativistic equation to find the radial velocity for the quasar, we obtain $v/c=(4.78^2-1)/(4.78^2+1)=...=0.92$.  In other words, this quasar appears to be receding from us at 92\% of the speed of light.''}
\xcite{21}{Taylor, E.~F. and Wheeler, J.~A. 1991, {\em Spacetime Physics: introduction to special relativity}, (New York: W.~H.~Freeman \& Co.) p.~264, Ex.~8-23}
\xcite{22}{Hu, Y., Turner, M.~S. and Weinberg, E.~J. 1993, \prd, 49(8), 3830--3836, {\em Dynamical solutions to the horizon and flatness problems}, ``\dots many viable implementations of inflation now exist.  All involve two key elements:  a period of superluminal expansion..." They define superluminal expansion later in the paper as ``Superluminal expansion might be most naturally defined as that where any two comoving points eventually lose causal contact.''  Their usage of superluminal conforms to this definition, and as long as the reader is familiar with this definition there is no problem.  Nevertheless, we should use this definition with caution because even if the recession velocity between two points is $\dot{D}>c$ this does not mean those points will eventually lose causal contact.}
\xcite{23}{Khoury, J., Ovrut, B.~A., Steinhardt, P.~J. and Turok, N. 2001, \prd, 64(12), 123522, {\em Ekpyrotic universe: Colliding branes and the origin of the hot big bang}, ``The central assumption of any inflationary model is that the universe underwent a period of superluminal expansion early in its history before settling into a radiation-dominated evolution.''  They evidently use a definition of `superluminal' that is common in inflationary discussions, but again, a definition that should be used with caution.}
\xcite{24}{Lovell, B. 1981, {\em Emerging Cosmology}, (New York: Columbia U. Press) p.~158, ``\dots observations with contemporary astronomical instruments transfer us to regions of the universe where the concept of distance loses meaning and significance, and the extent to which these penetrations reveal the past history of the universe becomes a matter of more sublime importance''}
\xcite{25}{McVittie, G.~C. 1974, \qjras, 15(1), 246--263, {\em Distance and large redshifts},``\dots conclusions derived from the assertion that this or that object is `moving with the speed of light' relative to an observer must be treated with caution.  They have meaning only if the distance and time used are first carefully defined and it is also demonstrated that the velocity so achieved has physical significance.''}
}

\section*{Acknowledgments}
We thank John Peacock and Geraint Lewis for pointing out an error in an earlier version of this paper.
We would also like to thank Jochen Liske, Hugh Murdoch, Tao Kiang, John Peacock, Brian Schmidt, Phillip Helbig, John Webb and Elizabeth Blackmore for useful discussions.  TMD is supported by an Australian Postgraduate Award. CHL acknowledges an Australian Research Council Research Fellowship.

\section*{References}






\reference Bennett, C.~L., Halpern, M., Hinshaw, G. \etal~2003, ApJS, 148, 1
\reference Chen, H.-W., Lanzetta, K.~M. and Pascarelle, S. 1999, Nature, 398, 586
\reference Davis, T.~M. and Lineweaver, C.~H. 2001, in Cosmology and Particle Physics 2000, ed. R.~Durrer, J.~Garcia-Bellido and M.~Shaposhnikov, (New York: American Institute of Physics conference proceedings, Volume 555), 348
\reference Davis, T.~M and Lineweaver, C.~H. 2004, in preparation
\reference Davis, T.~M., Lineweaver, C.~H. and Webb, J.~K. 2003, Am. J. Phys., 71, 358
\reference Ebert, R. and Tr\"umper, M. 1975, Ann. NY Acad. Sci., 262, 470
\reference Ellis, G.~F.~R. and Rothman, T. 1993, Am. J. Phys., 61, 883
\reference Fan, X., Strauss, M.~A., Schneider, D.~P. \etal~2003, AJ, 125, 1649
\reference Goldhaber, G., Dustua, S., Gabi, S. \etal~1997, in ed. R.~Canal, P.~Ruiz-LaPuente and J.~Isern, NATO ASIC Proc.~486: Thermonuclear Supernovae, 777
\reference Goldhaber, G., Groom, D.~E., Kim, A. \etal~2001, ApJ, 558, 359
\reference Gudmundsson, E.~H. and Bj\"ornsson, G. 2002, ApJ, 565, 1
\reference Harrison, E.~R. 1981, Cosmology: the science of the universe, 1st Ed. (Cambridge: Cambridge U. Press), and 2nd Ed. (2000)
\reference Harrison, E.~R. 1993, ApJ, 403, 28
\reference Kiang, T. 1991, Acta Astrophysica Sinica, 11, 197
\reference Kiang, T. 1997, Chinese Astronomy and Astrophysics, 21, 1
\reference Kiang, T. 2001, Chinese Astronomy and Astrophysics, 25, 141
\reference Lake, K. 1981, ApJ, 247, 17
\reference Landsberg, P.~T. and Evans, D.~A. 1977, Mathematical Cosmology, an introduction (Oxford: Clarendon Press)
\reference Leibundgut, B., Schommer, R., Phillips, M. \etal~1996, ApJ, 466, L21
\reference Liske, J. 2000, MNRAS, 319, 557
\reference Loeb, A. 1998, ApJ, 499, L111
\reference Murdoch, H.~S. 1977, QJRAS, 18, 242
\reference Murdoch, H.~S. 1993, Quasars - how far? how fast?, unpublished, Sect.~1
\reference Outram, P.~J., Chaffee, F.~H. and Carswell, R.~F. 1999, MNRAS, 310, 289
\reference Page, D.~N. 1993, unpublished, gr-qc/9303008
\reference Peacock, J.~A. 1999, Cosmological Physics, (Cambridge: Cambridge U. Press)
\reference Perlmutter, S., Aldering, G., Goldhaber, G. \etal~1999, ApJ, 517, 565
\reference Press, W.~H., Flannery, B.~P., Teukolsky, S.~A. and Vetterling, W.~T. 1989, Numerical Recipes (Fortran Version), (Cambridge: Cambridge U. Press)
\reference Riess, A.~G., Filippenko, A.~V., Challis, P. \etal~1998, AJ, 116, 1009
\reference Riess, A.~G., Filippenko, A.~V., Leonard, D. \etal~1997, AJ, 114, 722
\reference Rindler, W. 1956, MNRAS, 6, 662
\reference Rindler, W. 1977, Essential relativity, special, general and cosmological, (New York: Springer)
\reference Sandage, A. 1962, ApJ, 136, 319
\reference Silverman, A.~N. 1986, \ajp, 54, 1092
\reference Stuckey, W.~M., \ajp, 60, 142
\reference Weinberg, S. 1972, Gravitation and cosmology, (New York: Wiley)
\reference Wilson, O.~C. 1939, ApJ, 90, 634
\reference Wright, E. L. 2001, Errors in tired light cosmology, www.astro.ucla.edu/$\sim$wright/tiredlit.htm

\end{document}